\documentclass[12pt]{article}
\pdfoutput=1

\usepackage{cite}
\usepackage{booktabs}
\usepackage[english]{babel}
\usepackage{amsmath,amssymb,amsbsy,amstext, amsthm, simplewick}
\usepackage{hyperref}
\usepackage{graphicx}
\usepackage{amsfonts}
\usepackage{amssymb}
\usepackage[small]{caption}
\usepackage{upgreek}
\usepackage[svgnames,dvipsnames,x11names,table]{xcolor}
\usepackage{multirow} 
\usepackage{geometry}
\usepackage[hang,flushmargin]{footmisc}
\usepackage{bm}
\usepackage{braket}
\usepackage{subcaption}
\usepackage{mathtools}
\usepackage{setspace}
\usepackage{cleveref}
\usepackage{comment}
\usepackage{scalerel}
\usepackage[normalem]{ulem}
\usepackage{slashed}
\usepackage{siunitx}
\usepackage{aas_macros}

\hypersetup{
    colorlinks=true,
    linkcolor={red!50!black},
    citecolor={blue!50!black},
    urlcolor={blue!80!black}
}

\usepackage{colortbl}

\makeatletter
\newlength{\apb@width}
\newcommand{\autoparbox}[2][c]{\settowidth{\apb@width}{#2}\parbox[#1]{\apb@width}{#2}}

\makeatother

\definecolor{lightgray}{gray}{0.9}

\usepackage[framemethod=default]{mdframed}
\newmdenv[skipabove=7pt,
skipbelow=7pt,
rightline=false,
leftline=false,
topline=false,
bottomline=false,
backgroundcolor=gray!10,
linecolor=gray,
innerleftmargin=5pt,
innerrightmargin=5pt,
innertopmargin=5pt,
innerbottommargin=5pt,
leftmargin=0cm,
rightmargin=0cm,
linewidth=4pt]{eBox}

\usepackage[most]{tcolorbox}
\tcbset{colback=white, colframe=black,
        highlight math style= {enhanced, 
            colframe=red,colback=red!10!white,boxsep=0pt}
        }

\crefname{table}{Table}{Tables}
\crefname{equation}{Eq.}{Eqs.}
\crefname{appendix}{App.}{Apps.}
\crefname{section}{Sec.}{Secs.}
\crefname{figure}{Fig.}{Figs.}

\numberwithin{equation}{section}

\def\beq{\begin{equation}}
\def\eeq{\end{equation}}

\def\bea{\begin{eqnarray}}
\def\eea{\end{eqnarray}}

\def\dd{{\rm d}}

\def\Neff{N_{\rm eff}}
\def\beq{\begin{equation}}
\def\eeq{\end{equation}}
\def\bea{\begin{eqnarray}}
\def\eea{\end{eqnarray}}

\def\dd{{\rm d}}

\DeclareRobustCommand{\SkipTocEntry}[4]{}

\newcommand{\s}{\hspace{0.8pt}}

\definecolor{colorTC}{rgb}{.2,.7,.2}

\definecolor{blue3}{RGB}{31, 119, 180}
\definecolor{red3}{RGB}{	214, 39, 40}
\definecolor{orange3}{RGB}{255, 127, 14}
\definecolor{green3}{RGB}{44, 160, 44}

\begin{document}

\begin{titlepage}
\setcounter{page}{1} \baselineskip=15.5pt 
\thispagestyle{empty}
$\quad$
\vskip 50 pt

\begin{center}
{\fontsize{18}{18} \bf Cosmological Implications of a \\
\vskip 10pt
Neutrino Mass Detection}
\end{center}

\vskip 20pt
\begin{center}
\noindent
{\fontsize{12}{18}\selectfont   Daniel Green$^{\s 1}$ and Joel Meyers$^{\s 2}$}
\end{center}

\begin{center}
\vskip 4pt
\textit{ $^1${\small Department of Physics, University of California at San Diego,  La Jolla, CA 92093, USA}
}
\vskip 4pt
\textit{ $^2${\small Department of Physics, Southern Methodist University, Dallas, TX 75275, USA}}

\end{center}

\vspace{0.4cm}
 \begin{center}{\bf Abstract}
 \end{center}
\noindent  The next generation of cosmological surveys are expected to measure a non-zero sum of neutrino masses, even down to the minimum value of 58~meV inferred from neutrino flavor oscillation.  The implications of such a measurement for the physics of neutrinos have been well documented; in contrast, the cosmological implications of such a measurement have received less attention.  In this paper, we explore the impact of a neutrino mass detection consistent with $\sum m_\nu = 58$~meV for our understanding of the history and contents of the universe.  We focus primarily on three key areas: the thermal history of the universe, clustering of matter on diverse scales, and the application to dark matter and dark sectors.  First we show that a detection of non-zero neutrino mass would provide a unique connection between the cosmic neutrino background, which is detected gravitationally, and neutrinos measured on Earth.  We then discuss how the consistency of a detection between multiple probes will impact our knowledge of structure formation.  Finally, we show how these measurements can be interpreted as sub-percent level tests of dark sector physics.  

\end{titlepage}

\setcounter{page}{2}

\restoregeometry

\begin{spacing}{1.4}
\newpage
\setcounter{tocdepth}{2}
\tableofcontents
\end{spacing}

\setstretch{1.1}
\newpage

\section{Introduction}

The prospect that precision cosmological measurements could reveal the effects of non-zero neutrino masses has drawn broad attention from both the cosmology and high energy physics communities~\cite{Dolgov:2002wy,Lesgourgues:2006nd,Dvorkin:2019jgs}.  As neutrinos are the last of the Standard Model particles with unknown mass parameters~\cite{ParticleDataGroup:2020ssz}, a natural focus of the community has been on the implications for neutrino physics and the relation to other neutrino experiments.  Yet, a cosmological detection of $\sum m_\nu$ relies on a number of non-trivial events in cosmic history, most notably the creation of a cosmic neutrino background at early times and the growth of structure at late times.  It is therefore natural to consider how the detection of the subtle cosmological influence of massive neutrinos informs our understanding of cosmology, independent of its implications for the physics of the neutrinos themselves.

The non-zero mass of cosmic neutrinos is measurable in cosmology because it suppresses the growth of structure on small scales.  Although the neutrinos are non-relativistic and redshift like matter in the late universe, their velocity remains sufficiently large to prevent neutrinos from falling into the gravitational wells created by dark matter on scales smaller than their effective Jeans scales.  Since the neutrinos do contribute to the total matter density as measured by the expansion of the universe, probes of the density fluctuation on scales smaller than the Jeans scale will see a smaller than expected amplitude of clustering~\cite{Lesgourgues:2006nd,Wong:2011ip,Lesgourgues:2012uu,Lattanzi:2017ubx}.  This includes the measurements of gravitational lensing of the cosmic microwave background (CMB)~\cite{Kaplinghat:2003bh}, clustering and weak lensing of galaxies~\cite{Hu:1997mj,Cooray:1999rv}, and the number density of galaxy clusters~\cite{Abazajian:2011dt,Carbone:2011by}.  Since all these probes are unable to resolve the Jeans scale itself, a measurement of this suppression is limited by our knowledge of the primordial scalar amplitude ($A_s$) from the CMB.  Our ability to measure $A_s$ is in turn limited by the uncertainty on the optical depth to reionization ($\tau$), since the observed power of CMB fluctuations on all but the largest angular scales is proportional to $A_s \exp(-2\tau)$~\cite{Zaldarriaga:1996ke,Hu:1999vq}.  Precise measurement of the reionization bump in the CMB polarization power spectrum on large scales, which is proportional to $\tau^2$, would decrease our uncertainty on $\tau$ and thus allow for improved cosmological measurements of neutrino mass.

Given the broad interest in the properties of neutrinos, there has been much attention paid to reaching the threshold where a cosmological detection of the effects of neutrino mass will be possible.
However, this enterprise involves combining a variety of cosmological measurements to break the degeneracies described above.  As a result, surveys with broadly different characteristics report similar forecasts for $\sum m_\nu$ because all of the cosmological measurements are limited by the same external data (most notably the uncertainty on $\tau$). 

From a cosmological perspective, there is much reason to be interested in the detection and measurement of $\sum m_\nu$ in future surveys.  Cosmic neutrinos are the earliest known relic of the hot Big Bang, with decoupling starting around one second after the Big Bang, at temperatures of order an MeV.  Their temperature today is expected to be 1.95~K, which is smaller than the CMB temperature primarily due to the conversion of entropy in electrons and positrons into photons. During the radiation era, the neutrinos make up 41\% of the energy density in the universe, leading to large gravitational effects~\cite{Bashinsky:2003tk,Baumann:2015rya,Baumann:2017lmt,Green:2020fjb} that are imprinted on the CMB~\cite{Follin:2015hya,Baumann:2015rya,Brust:2017nmv,Planck:2018vyg}, primordial abundances of light elements~\cite{Cyburt:2015mya}, and baryon acoustic oscillations (BAO)~\cite{Baumann:2019keh} (via their contribution to the radiation density, $\Neff$).  While the detection of these effects gives us confidence that we have observed the cosmic neutrino background, the gravitational nature of the signature leaves no direct connection to the particles we observe in the lab.  In principle, the detection of a non-zero neutrino mass in cosmological observations provides the opportunity to relate the lab and cosmic neutrinos.  Furthermore, as the signal itself depends on the pristine state of neutrinos created in the early universe, it provides a sensitive probe of new physics in the neutrino sector.

More generally, the cosmological neutrino mass measurement is a comparison of the small scale power across a wide range of redshifts, scales, and tracers.  Viewed only as independent measurements of $\sum m_\nu$, better measurements of the matter power spectrum offer little additional constraining power. Yet, a suppression of power can arise on a variety of scales due to dark sector interactions and from other massive relics, and additional measurements can offer a window into such processes.  In addition, the nonlinear physics of structure formation, including the role of baryons, impacts weak lensing, galaxy clustering, and cluster counts in distinct ways. In contrast, CMB lensing measurements are largely insensitive to short distance physics.  As a result, the combination of these probes will allow us to disentangle dark sector and neutrino physics from structure formation, shedding light on both.  

This paper is organized as follows: Section~\ref{sec:CnuB} reviews of the physics of the cosmic neutrino background and its observable effects, Section~\ref{sec:clustering} explores how various probes of large scale structure are sensitive to the matter power spectrum and its suppression by $\sum m_\nu$, the implications of a cosmological measurement of $\sum m_\nu$ for dark sector physics are discussed in Section~\ref{sec:dark}, and we conclude in Section~\ref{sec:Discussion}.

\section{Cosmic Neutrino Background}
\label{sec:CnuB}

In this section, we discuss the implications of the neutrino mass measurement for the physics of the cosmic neutrino background and early universe physics.  In particular, the cosmological signature of neutrino mass is dependent on the neutrino temperature and abundance.  A cosmological detection of the neutrino mass provides a test of the entire cosmic history of these neutrinos.  Furthermore, it ties together cosmological and lab-based measurements.  

\subsection{The Physics of a Neutrino Mass Detection}
\label{subsec:physics_of_mnu}

In order to understand the impact of a detection of neutrino mass, we must first understand the physics responsible for its observable cosmological signatures.  Given current constraints on the sum of neutrino masses, $\sum m_\nu < 120$ meV (95\%)~\cite{Planck:2018vyg}, we can infer that the cosmic neutrinos became non-relativistic after recombination, and therefore we can focus on the impact of neutrino mass on the matter power spectrum.

For a standard cosmology, the neutrino temperature today is $1.95$~Kelvin, which is slightly lower than the CMB temperature, since $e^+ e^-$ annihilation occurs after weak decoupling of the neutrinos begins.  Given this temperature, a 50~meV neutrino would become non-relativistic at redshift $z\approx 100$, when the average neutrino momentum, $p_\nu \approx 3 T_\nu$, drops below its mass $p_\nu < m_\nu$.  This occurs long before any direct observational probes of the large scale structure of the universe while also being much later than recombination.  In the non-relativistic regime, neutrinos contribute to the expansion like the dark matter and baryons, with a total energy density given by
\beq
\Omega_\nu h^2 =  6.2 \times 10^{-4} \, \left( \frac{\sum m_\nu}{58 \, {\rm meV}} \right) \ .
\eeq
Given the current measurement of $\Omega_\mathrm{m}  h^2 = 0.1428 \pm 0.0011$~\cite{Planck:2018vyg}, we see that neutrinos contribute only $0.4-1$ percent of the matter density and therefore make a small (but important) impact on the expansion rate at late times.  

\begin{figure}[ht!]
    \centering
    \includegraphics[width=5in]{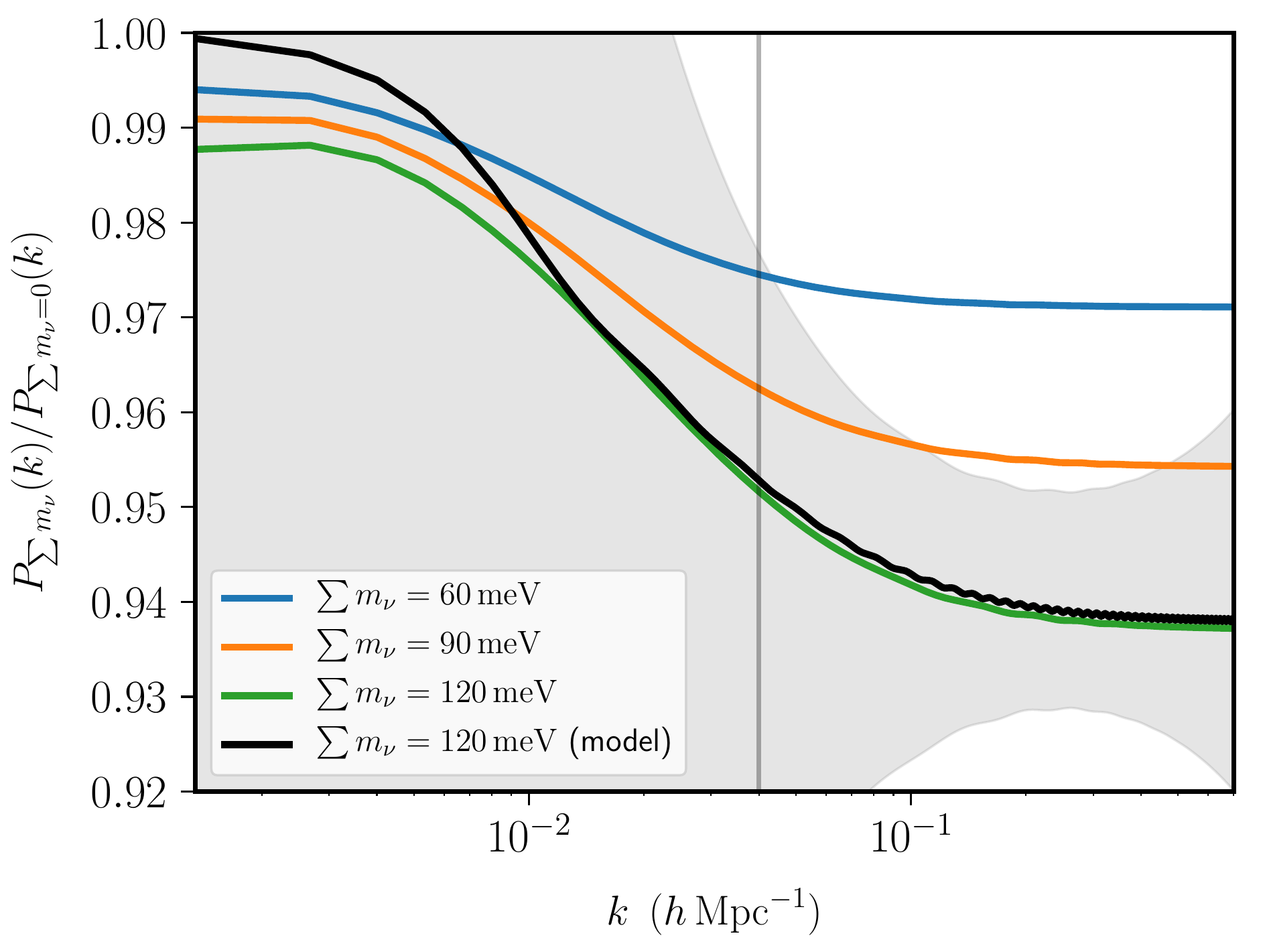}
    \caption{Effect of neutrino mass on the matter power spectrum at $z=0$.  The colored lines indicate the results from  \texttt{CLASS}~\cite{Blas:2011rf} when varying $\sum m_\nu$ while holding fixed $H_0$, $\Omega_\mathrm{c} h^2$, $\Omega_\mathrm{b} h^2$, and $A_s$. For comparison, the vertical line indicates the estimated free-streaming scale from Equation~(\ref{eq:kfs_nu}) and the solid black line shows the result of the solution to the fluid equations for $\sum m_\nu = 120$~meV, showing excellent agreement between the fluid approximation and the Boltzmann code.  The grey band indicates the $1\sigma$ uncertainty associated with a BOSS-like galaxy survey~\cite{Font-Ribera:2013rwa} including both cosmic variance and shot noise, which dominate at low and high $k$ respectively.
    }
    \label{fig:mnu}
\end{figure}

Although the massive neutrinos contribute to the expansion at late times like any other form of non-relativistivc matter, their thermal velocities are not negligible like those of the baryons and dark matter.  Instead, neutrinos move over cosmological distances in a Hubble time and thus their fluctuations do not evolve like cold matter.  The overdensities of cold matter (the baryons and cold dark matter) $\delta_\mathrm{cb}= \frac{\delta\rho_\mathrm{c} + \delta\rho_\mathrm{b}}{\bar{\rho}_\mathrm{c}+\bar{\rho}_\mathrm{b}}$ and the neutrinos $\delta_\nu = \frac{\delta \rho_\nu}{\bar{\rho}_\nu}$ evolve as a coupled system of fluids according to
\begin{align}
    \ddot{\delta}_{\rm cb}+\frac{4}{3 t} \dot{\delta}_{\rm cb}&=\frac{2}{3 t^{2}}\left[f_\nu \delta_{\nu}+(1-f_\nu) \delta_{\rm cb}\right] \label{eq:nu_cmd1}\\
    \ddot{\delta}_{\nu}+\frac{4}{3 t} \dot{\delta}_{\nu}&=-\frac{2 \alpha}{3 t^{2}} \delta_{\nu}+\frac{2}{3 t^{2}}\left[f_\nu \delta_{\nu}+(1-f_\nu) \delta_{\rm cb}\right] \label{eq:nu_cmd2}
\end{align}
where
\beq
    \alpha \equiv \frac{3 k^{2} c_\nu^{2} t^{2}}{2 a^{2}}=\frac{k^2}{k_{\rm fs}^2} \ , 
    \quad c_\nu \equiv \frac{\langle p_\nu \rangle}{m_\nu}  \ , 
    \quad f_\nu \equiv \frac{\Omega_{\nu}}{\Omega_\mathrm{m}}  \ .
\eeq
This phenomenon is usually understood in terms of the scale on which neutrinos free-stream, expressed as a comoving wavenumber as
\beq
    k_{\rm fs} = 0.04 \, h \, {\rm Mpc}^{-1} \times \frac{1}{1+z} \, \left(\frac{\sum m_\nu}{58 \, {\rm meV}}\right) \, . 
    \label{eq:kfs_nu}
\eeq
For $k \gg k_{\rm fs}$, the neutrinos do not cluster, since they behave as if they are subject to a Jeans scale determined by $k_{\rm fs}$.  Overdensities in dark matter and baryons on scales $k\gg k_{\rm fs}$ will continue to grow, but now in a universe where the rate of expansion includes the effects of neutrinos while the local force of gravity needed for clustering does not.  This can be seen from the $\alpha \gg 1$ limit of Equation~(\ref{eq:nu_cmd2}) which drives $\delta_\nu \to 0$. The most direct consequence is that the amplitude of linear fluctuations 
$\delta_\mathrm{m} = f_\nu \delta_\nu +(1-f_\nu)\delta_{\rm cb}$ 
only gets contributions from the dark matter and baryons.  A more subtle consequence is that the clustering of neutrinos reduces the rate of growth of the dark matter and baryons due to the factor of %
$(1-f_\nu)$
in Equation~(\ref{eq:nu_cmd1}).  Combining these two effects, on these scales the power spectrum is suppressed by 
\beq
    P_{\sum m_\nu} (k\gg k_{\rm fs}, z ) \approx  \left(1 - 2 f_\nu  -\frac{6 }{5} f_\nu\log \frac{1+z_\nu}{1+z} \right) P_{\sum m_\nu =0}(k\gg k_{\rm fs}, z ) \label{eq:suppression}
\eeq
where $z_\nu$ is the redshift when the neutrinos become non-relativistic 
($\langle p_\nu \rangle = m_\nu$).
Notice that this formula depends both on the $\sum m_\nu$ via the $f_\nu$ and also on the temperature of neutrinos via the duration of the non-relativistic regime $\log \frac{1+z_\nu}{1+z}$.  The enhancement by the $\log \frac{1+z_\nu}{1+z}$ is due to the change to the growth rate and is responsible for the larger than expected suppression of power due to massive neutrinos.  These effects are shown in Figure~\ref{fig:mnu} using both the full solution from \texttt{CLASS}~\cite{Blas:2011rf} and from solving the fluid equations, Equations~\eqref{eq:nu_cmd1} and \eqref{eq:nu_cmd2}.

In practice, the scale $k_{\rm fs}$ is too small\footnote{In fact, because $k_{\rm fs}$ decreases with increasing $z$, $k_{\rm fs}(z=0)$ overestimates the $k$ where the scale dependence would be observable. See Figure~\ref{fig:mnu}.} to be observable in near term CMB or large scale structure surveys.  As a consequence, the measurement of neutrino mass is effectively a comparison of the amplitude of $P (k > k_{\rm fs},z)$ to the primordial amplitude $A_s$ as determined by the primary CMB spectrum.  As a result, measurements of $\sum m_\nu$ are subject to strong degeneracies with any parameter that limits our knowledge of either $A_s$ or the amplitude of the matter power spectrum $P(k,z)$.  

\subsection{Implications for Cosmic Neutrinos}


The existence of the cosmic neutrino background is a highly non-trivial and fascinating aspect of the standard Big Bang cosmology.  As neutrino decoupling begins prior to nucleosynthesis, it is the earliest direct probe of the universe in the Standard Model (unless we observe primordial gravitational waves).  

While direct detection of cosmic neutrinos remains elusive, cosmology has provided an abundance of evidence for their existence.  Most straightforwardly, neutrinos contribute 41\% of the energy density during the radiation era and thus significantly impact the expansion rate in the early universe.  Measurement of $\Neff$ via primordial abundances and the damping tail of the CMB confirm the need for such a component of radiation at 30$\sigma$~\cite{Cyburt:2015mya,Planck:2018vyg}.  In addition, measurements of the structure of acoustic peaks in the CMB and BAO have shown that this component of radiation is free streaming (rather than a fluid) at 10$\sigma$ and 2.5$\sigma$, respectively~\cite{Follin:2015hya,Baumann:2015rya,Baumann:2019keh}.  In all cases, the evidence for the existence of cosmic neutrinos is being provided by their gravitational influence in the relativistic regime, where they are effectively massless.

Naturally, one might hope to directly connect the free-streaming radiation relevant for cosmology to the particles we observe in the lab.  In principle, a new light particle species carrying approximately the same energy density as neutrinos would also fit observations.  The apparent coincidences between the observed energy could be accomplished by a decay or annihilation channel for the neutrinos that would leave a similar energy density in dark sector radiation.  Alternatively, one could imagine mirror sectors with similar physical laws where the majority of $\Neff$ is described by some mirror neutrino-like particle (this possibility may even help explain the Hubble tension~\cite{Cyr-Racine:2021alc}).  While such scenarios often leave additional cosmological signatures, we emphasize that cosmological signatures of $\Neff$ are no different for cosmic neutrinos than for a free-streaming particle species in the dark sector. 

A detection of neutrino mass with $\sum m_\nu =  58$~meV presents an opportunity to directly connect the particle we observe in the lab to an important source of gravity both in the late and early universe.  From the measurement of neutrino oscillations, we know the squared mass splittings between the mass eigenstates but not the absolute mass scale~\cite{deSalas:2017kay}.  In this sense, the cosmological observable is not directly related to the lab.  However, in the case where $\sum m_\nu \approx 60$ meV, we know that the masses in the neutrino sector are hierarchical such that $m_1 \ll m_3$ and that $\sum m_\nu \approx m_{3} \approx \sqrt{|\Delta m_{31}^2|}$.  In this case, the agreement between the cosmological measurement and the minimum mass would therefore be a confirmation that the heaviest neutrino seen in the lab is the same particle responsible for the gravitational influences seen in cosmological observables (assuming there is no fine tuning to make the neutrino masses identical to the mass of some other particle).

Naturally, we should ask with what significance we need to measure $\sum m_\nu$ before we could make such an inference.  If our expectation is that particle masses are non-degenerate, then a simple qualitative estimate is that we confirm that the masses are of the correct order of magnitude, namely 
\beq
100 \ {\rm meV}  >\sum m_\nu >  10 \ {\rm meV} \ .
\eeq
Reaching this sensitivity would require $\sigma(\sum m_\nu) < 20 \ {\rm meV}$. Interestingly, this kind of sensitivity is not possible without an improved measurement of $\tau$ and thus motivates efforts to reach the cosmic variance limit using CMB observations from the ground~\cite{Harrington:2016jrz}, balloons~\cite{BFORE:2017taq,Masi:2019wqa}, or space~\cite{Hazumi:2019lys,Hanany:2019lle}.  In addition, there exist proposals to determine $\tau$ using 21~cm intensity mapping~\cite{Liu:2015txa,Alvarez:2019pss} or the kinetic Sunyaev-Zel'dovich trispectrum~\cite{Smith:2016lnt, Ferraro:2018izc,Alvarez:2020gvl}.

This qualitative estimate of the required sensitivity is consistent with expectations from models where the energy density in hot relics does not include contributions from all the neutrino species seen in the lab.  Concretely, if some Standard Model neutrinos decay to a dark sector in the early universe, the largest residual $\sum m_\nu$ arises in a scenario where $\nu_3 \to \nu_2 + \phi$ where $\phi$ is some new light particle with $m_\phi \ll m_2$, such that $\sum m_\nu \approx 17$~meV.
We can exclude such a possibility at 95\% confidence if we detect $\sum m_\nu\approx 60$~meV with $\sigma(\sum m_\nu) < 20$~meV.

Interestingly, the neutrino lifetime can also be constrained from cosmological observables as discussed in~\cite{Chacko:2019nej,Chacko:2020hmh}.  In cases where the decay is rate is large enough to reduce $\sum m_\nu \ll 60$ meV, the redshift dependence of the neutrino density is also potentially observable.  However, this is only assuming the simplest model of a constant decay rate.  One could construct scenarios where the heaviest neutrinos are annihilated or decay through a process that is more sensitive to the temperature of the universe and thus could evade constraints from the early and late universe.  Examples of such time-dependent effects arise naturally with light right-handed neutrinos and/or interactions with dark sectors~\cite{Green:2021gdc}.


\subsection{Implications for the History of the Universe}

The measurement of $\Neff$ from the primordial abundance of light nuclides establishes the existence of the cosmic neutrino background a few minutes after the Big Bang, primarily through its influence on the expansion rate~\cite{Cyburt:2015mya}.  Measurements of $\Neff$ with the CMB and BAO further establish that this background persisted \num{380000} years after the Big Bang during the era of recombination.   

The detection of $\sum m_\nu =58$~meV in CMB lensing, the clustering and lensing of galaxies, or cluster abundances would confirm the presence of neutrinos at low redshifts, $z=0$\textendash{}$10$, when the universe is billions of years old.  However, as seen in Equation~\eqref{eq:suppression}, the observed amplitude is also sensitive to the time at which neutrinos became non-relativistic, namely $3 T_\nu \approx 50$~meV or $z \approx 100$.  This is a remarkable statement because it breaks the degeneracy between the total energy in neutrinos (light relics) and the temperature of the neutrinos.  Concretely, since
\beq
    \Neff =  N_\nu \, \left( \frac{11}{4} \right)^{4/3} \left(\frac{T_\nu}{T_\gamma}\right)^4 \, ,
\eeq
where $N_\nu$ is the number of neutrino species, we can change $T_\nu$ holding $\rho_\nu$ fixed and leave cosmological observables unchanged.  For example, we could lower the temperature of the neutrinos but add new species.  However, as discussed in Section~\ref{subsec:physics_of_mnu}, the amplitude of the suppression of power is much larger than one would expect from just the energy density in neutrinos today because of the high redshift where the heaviest neutrino becomes non-relativistic.
 
The universe at $z \approx 100$ is otherwise unobservable with near-term observations.  The CMB is only very weakly sensitive to changes in that period\footnote{The CMB is sensitive to injection of electrons at that time due to re-scattering of the CMB photons.  Dark matter annihilation~\cite{Slatyer:2017sev,Green:2018pmd} or decay~\cite{Kaplinghat:1999xy,Peter:2010au} provide examples of such a possibility.} as the integrated Sachs-Wolfe effect and gravitational lensing get negligible contributions from that epoch.  The detection of $\sum m_\nu = 58$ meV would constrain the possibility of significant heating or cooling of neutrinos during that era.  Dark matter and/or dark radiation interactions provide obvious mechanisms to change the neutrino distribution after recombination, particularly for light right-handed neutrinos~\cite{Green:2021gdc}.
 
Substantial cooling of the neutrinos would also increase the wavenumber of the free-streaming scale to a potentially observable range. 
Specifically, 
\beq
    k_{\rm fs} = \sqrt{\frac{3}{2}}\frac{aH}{c_\nu} 
\eeq
so using 
\beq
    c_\nu = \frac{\langle p_\nu \rangle}{m_\nu} = \frac{3 T_\nu}{m} \approx 8.7 \times 10^{-3} \left(\frac{58 \, {\rm meV}}{\sum m_\nu}\right)\, (1+z) \, c   \ ,
\eeq
we have
\beq
    k_{\rm fs}(z) = 0.04 \, h \, {\rm Mpc}^{-1}  \, \left(\frac{\sum m_\nu}{58 \, {\rm meV}}\right) \times \left(\frac{1.95 \, {\rm K}}{T_\nu(z=0)}\frac{1}{1+z}\right)  \ .
    \label{eq:kfs_nu_cooling}
\eeq
If $k_{\rm fs} \approx 0.1 \, h \, {\rm Mpc}^{-1}$ the scale dependence would be visible with high signal to noise in a galaxy survey like BOSS~\cite{BOSS:2016hvq}, DESI~\cite{DESI:2016fyo}, or LSST~\cite{LSSTScience:2009jmu}.  This will put stringent limits on sources of neutrino cooling.  In contrast, heating the neutrinos would decrease $k_{\rm fs}$ and thus it would remain unobservable.  Nevertheless, heating would decrease $z_\nu$
and thus would decrease the suppression of the matter power spectrum, effectively acting as $\sum m_\nu < 58$~meV.  These changes to free streaming also arise in many models of dark sectors, which we will expand upon in Section~\ref{sec:dark}.



\section{Clustering and Structure Formation}
\label{sec:clustering}

In this section we discuss some of the cosmological observables that can be used to measure the suppression of small scale power that results from  non-zero $\sum m_\nu$.  We will focus on CMB lensing, angular power spectra of galaxy density, and number counts of galaxy clusters and show the range of wavenumber and redshift to which each observable is sensitive.

Several observables of cosmological structure are most effectively described in terms of line-of-sight projections of random fields.
Much of the statistical information contained in the distribution of matter throughout space can be gleaned from the angular power spectra of these fields.
On all but the largest angular scales, these power spectra can be simplified by use of the Limber approximation~\cite{1953ApJ...117..134L,Kaiser:1991qi,Kaiser:1996tp,LoVerde:2008re}.
The angular power spectrum for large scale structure fields $X$ and $Y$ can be expressed in the Limber approximation as
\begin{align}
    C_\ell^{XY} = \int \dd{z} \frac{H(z)}{\chi^2(z)} W^{X}(z) W^{Y}(z) P \left( k= \frac{\ell}{\chi(z)},z \right) \, ,
    \label{eq:Limber}
\end{align}
where $H(z)$ is the Hubble rate at redshift $z$, $\chi(z)$ is the comoving distance to redshift $z$, and $W^{X}(z)$ is the redshift kernel for field $X$.  Each observable we discuss is sensitive to the matter power spectrum in different regimes due to the differing redshift kernels.

The goal is to investigate the complimentarity of these observables in the era of a neutrino mass detection. While they all produce similar forecasts in raw sensitivity to $\sum m_\nu$, these probes vary widely in their sensitivity to distance scales and redshifts, as we will see in Figure~\ref{fig:Pk_stagger}.  A consistent measurement of $\sum m_\nu$ across these different observables requires control over a variety of short distance physical effects.

\subsection{Observables}
\subsubsection{CMB Lensing}
\label{subsec:CMBLensing}

Gravitational lensing of the CMB results from the deflection of CMB photons due to gradients of the gravitational potential along our line of sight; see Ref.~\cite{Lewis:2006fu} for a review.
CMB lensing alters the observed temperature and polarization power spectra by broadening acoustic peaks, transferring power from large angular scales to small scales, and by converting $E$-mode polarization to $B$-mode polarization.
Lensing also alters the statistics of the CMB, inducing correlations between fluctuations of different wavenumber.
This change to the statistics can be used to reconstruct a map of the lensing deflection~\cite{Hu:2001kj}; see Appendix~\ref{app:LensingRecon}.
The large scale structure of the universe is the source of gravitational lensing, and so by measuring CMB lensing deflection, the CMB can be used to study the distribution of matter intervening between the surface of last scattering and our telescopes.

It is convenient to express the effects of CMB lensing in terms of a lensing potential 
\begin{align}
    \phi(\hat{\mathbf{n}}) \equiv -2 \int_0^{\chi_\star} \dd{\chi} \frac{\chi_\star-\chi}{\chi_\star \chi} \Psi(\chi \hat{\mathbf{n}}, \eta_0-\chi) \, ,
    \label{eq:LensingPotential}
\end{align}
where $\hat{\mathbf{n}}$ is the line of sight direction,  $\chi_\star$ is the comoving distance to the CMB last scattering surface at $z_\star \simeq 1090$, $\Psi(\mathbf{x},\eta)$ is the Weyl potential, and $\eta_0$ is the conformal time today.  
Here we have assumed vanishing spatial curvature.
The lensing deflection angle is the angular gradient of the lensing potential $\boldsymbol{\alpha}=\nabla_{\hat{\mathbf{n}}} \phi(\hat{\mathbf{n}})$, and it is also convenient to define the lensing convergence as $\kappa(\hat{\mathbf{n}}) = -\nabla_{\hat{\mathbf{n}}}^2 \phi(\hat{\mathbf{n}})/2$.

The CMB lensing potential is a projection of the integrated matter density over the entire line of sight from the surface of last scattering to our telescopes.
The redshift kernel for the CMB lensing convergence is
\begin{align}
    W^{\kappa}(z) = \frac{3}{2}\Omega_m \frac{H_0^2}{H(z)} (1+z) \chi(z) \frac{\chi_\star - \chi(z)}{\chi_\star} \, .
    \label{eq:LensingKernel}
\end{align}
The total matter overdensity including non-relativistic neutrinos $\delta_\mathrm{m}$ is used to compute the lensing power spectrum.  Therefore, in this notation, we have
\begin{align}
    \kappa(\hat{\mathbf{n}}) = \int \dd z \, W^\kappa(z) \delta_\mathrm{m}(\chi(z)\hat{\mathbf{n}},z) \, .
\end{align}

Data from the Planck satellite has been used to detect the effects of CMB lensing at 40$\sigma$~\cite{Planck:2018lbu}.
Upcoming CMB surveys, such as those from Simons Observatory~\cite{Ade:2018sbj}, CMB-S4~\cite{Abazajian:2016yjj}, PICO~\cite{Hanany:2019lle}, and CMB-HD~\cite{Sehgal:2019ewc}, will provide very low noise measurements of CMB temperature and polarization fluctuations, thereby enabling high fidelity lensing reconstruction.

We use the methods described in Appendix~\ref{app:LensingRecon} to forecast the lensing reconstruction noise for CMB-S4.
The reconstruction is computed using the minimum variance quadratic estimator~\cite{Hu:2001kj} including the improvement from iterative $EB$ reconstruction~\cite{Hirata:2003ka,Smith:2010gu}, assuming an instrument design consistent with Ref.~\cite{Abazajian:2019eic} taken to cover $f_\mathrm{sky}=0.5$.
The results are shown in Figures~\ref{fig:Pk_stagger} and \ref{fig:lensing_mnu}.

\subsubsection{Galaxy Surveys}
\label{subsec:GalaxyDensity}

Galaxy surveys act as a useful probe of the large scale structure of the universe.  These surveys do not directly measure the distribution of matter; instead they measure the distribution of galaxies, which are highly nonlinear objects with a complicated formation history.  The observed distribution of galaxies serves as a biased tracer of the underlying distribution of matter throughout the universe.  Since structure formation on large scales is determined entirely by gravitation, the physics of galaxy formation relevant for these scales can be captured by a set of bias parameters.  Marginalizing over these parameters allows for the robust extraction of large scale cosmological information from galaxy surveys~\cite{Desjacques:2016bnm}.

The cosmological constraining power of a galaxy survey is driven by the number density of observed objects, since the shot noise is given by the inverse of the comoving number density of galaxies in the survey.  
Galaxy surveys like the Vera C.~Rubin Observatory Legacy Survey of Space and Time (LSST) will catalog the positions and photometric redshifts of billions of galaxies~\cite{LSSTScience:2009jmu}.

For photometric galaxy surveys it is useful to analyze the angular power spectra of galaxy densities divided into redshift bins (see e.g.~\cite{Font-Ribera:2013rwa,Schmittfull:2017ffw,Yu:2018tem,Yu:2021vce}).  These power spectra can be calculated using the Limber approximation shown in Equation~\eqref{eq:Limber}.
The redshift kernel for the galaxy density in redshift bin $i$ is
\begin{align}
    W^{g_i}(z) = \frac{b_i(z)  \frac{\dd n_i}{\dd z}}{\left( \int \dd{z'} \frac{\dd n_i}{\dd z'} \right)} \ ,
    \label{eq:GalaxyKernel}
\end{align}
where $b_i(z)$ is the linear galaxy bias in redshift bin $i$, which we take to be given by $b(z)=1+z$.  Imaging surveys also make use of galaxy lensing, which has a redshift kernel very similar to that of CMB lensing
\begin{align}
    W^{\kappa_{gi}}(z) = \frac{3}{2}\Omega_m \frac{H_0^2}{H(z)} (1+z) \chi(z) \int_{z}^{\infty} \dd z_\mathrm{s} \,  n_{\mathrm{s}i}(z_\mathrm{s}) \frac{\chi(z_\mathrm{s}) - \chi(z)}{\chi(z_\mathrm{s})} \, ,
    \label{eq:GalaxyLensingKernel}
\end{align}
where $z_\mathrm{s}$ is the redshift of source galaxies and $n_{\mathrm{s}i}(z_\mathrm{s})$ is the redshift selection function of source galaxies for redshift bin $i$~\cite{Hu:2003pt,LSSTScience:2009jmu,Font-Ribera:2013rwa}.
The galaxy density is sensitive to the overdensity of cold dark matter and baryons $\delta_\mathrm{cb}$ (i.e. the matter density excluding neutrinos) while the lensing is due to the total matter overdensity including neutrinos $\delta_\mathrm{m}$.
We therefore have
\begin{align}
    g_i(\hat{\mathbf{n}}) &= \int \dd z \, W^{g_i}(z) \delta_\mathrm{cb}(\chi(z)\hat{\mathbf{n}},z) \nonumber \\
    \kappa_{gi}(\hat{\mathbf{n}}) &= \int \dd z \, W^{\kappa_{gi}}(z) \delta_\mathrm{m}(\chi(z)\hat{\mathbf{n}},z) \, .
\end{align}

The LSST $i<25$ Gold sample is modeled with a galaxy redshift distribution of the form $\frac{\dd n}{\dd z} \propto \left(\frac{1}{2z_0}\right) \left(\frac{z}{z_0}\right)^2 e^{-z/z_0}$, with $z_0 = 0.3$ and a total angular number density $\bar{n}_\mathrm{tot} = 40$~arcmin$^{-2}$ and is taken to cover $f_\mathrm{sky} = 0.4$~\cite{LSSTScience:2009jmu}.  The galaxy density shot noise power for redshift bin $i$ is given by the inverse of the angular number density of galaxies in that bin $N_\ell^{g_ig_i} = 1/\bar{n}_i$, and the galaxy lensing noise $N_\ell^{\kappa_{gi}\kappa_{gi}} = 0.3^2/(2\bar{n}_i)$ accounts for the intrinsic shape noise of galaxies~\cite{LSSTScience:2009jmu,Hearin:2011bp,Font-Ribera:2013rwa}.

\subsubsection{Galaxy Clusters}
\label{subsec:Clusters}

Clusters of galaxies are the most massive gravitationally bound objects in the universe.  
Galaxy clusters are very rare objects, forming at the highest peaks of the density field.  The cluster abundance and its evolution with redshift is a sensitive probe of cosmology~\cite{Carlstrom:2002na,Allen:2011zs}, especially of the matter power spectrum on small scales.  Clusters are also biased tracers of the underlying matter distribution with a bias much larger than that of individual galaxies, and their clustering can be used to probe the power spectrum on larger scales~\cite{Carbone:2011by}.

Galaxy clusters are observed across a wide range of the electromagnetic spectrum~\cite{Allen:2011zs}.  The hot gas present in galaxy clusters makes them shine brightly in X-rays.  The high density of stars makes clusters visible in the optical and infrared.  Clusters also leave an imprint on the CMB through the Sunyaev-Zel'dovich (SZ) effect~\cite{Zeldovich:1969ff,Sunyaev:1970er,Sunyaev:1980vz}.  The thermal SZ effect, whereby CMB photons undergo inverse Compton scattering with the hot electrons present in clusters, leaves a spectral distortion in the CMB in the direction of clusters.  The thermal SZ effect is insensitive to the redshift of the cluster, and upcoming CMB surveys will allow for the detection of tens of thousands of clusters including many at high redshift~\cite{Carlstrom:2002na,Abazajian:2016yjj}.
Along with cluster mass measurements (made possible with scaling relations and gravitational lensing measurements), these cluster catalogs can be used to probe the mass function, which is sensitive to the growth of structure on small scales.

Galaxy cluster number counts are computed from the number density of dark matter halos
\begin{align}
    \frac{\dd n(M,z)}{\dd M \dd z} = f(\sigma,z) \frac{\rho_\mathrm{cb}}{M} \frac{\dd \log \sigma^{-1}(M,z)}{\dd M} \, ,
    \label{eq:HMF}
\end{align}
with $\sigma(M,z)$ defined as
\begin{align}
    \sigma(M,z) = \frac{1}{2\pi^2} \int_{0}^{\infty} \dd k \, k^2 P(k,z) W^2(k,R)  \, ,
    \label{eq:sigmaMz}
\end{align}
where  $W(k,R)$ is the Fourier transform of the spherical top-hat window function of radius $R$, with the halo mass $M$ given by $M=\frac{4\pi \rho_\mathrm{cb}} {R^3}$~\cite{Carbone:2011by,Costanzi:2013bha}.  We use the $\texttt{hmf}$ code~\cite{Murray:2013qza,2014ascl.soft12006M} which implements the Tinker fit to the functional form of $f(\sigma)$ given by~\cite{Warren:2005ey}
\begin{align}
    f(\sigma) = A\left[\left(\frac{\sigma}{b}\right)^{-a} + 1 \right] e^{-c/\sigma^2} \, ,
    \label{eq:Tinkerf}
\end{align}
where $A$, $a$, $b$, and $c$ are the Tinker best-fit parameters~\cite{Tinker:2008ff}.  The number of galaxy clusters exceeding a given mass in each redshift bin $N_i$ (where $\dd N_i = n_i \dd V$) follows a Poisson distribution with variance given by $N_i$.

\subsection{Nonlinear Effects}
\label{subsec:NonLin}

While the density field on all scales at early times and on large scales at late times can be adequately described by linear perturbation theory, the nonlinear growth of structure significantly impacts the matter distribution on small scales at late times.  Perturbation theory can be utilized to provide reliable results in the mildly nonlinear regime~\cite{Bernardeau:2001qr}, though large-scale cosmological simulations are necessary when nonlinearities become more prominent.  The analytic halo model~\cite{Peacock:2000qk,Seljak:2000gq,Cooray:2002dia} provides a good qualitative description of the nonlinear regime, but it needs to be calibrated to simulations in order to provide  the precision necessary to compare with current and future observations.  The nonlinear matter power spectrum can be computed using halo model-inspired fits to numerical simulations, such as with the \texttt{HALOFIT} software~\cite{Smith:2002dz,Takahashi:2012em,Bird:2011rb}.

In addition to purely gravitational nonlinearities in the growth of structure, the distribution of matter on small scales is also impacted by baryonic effects including supernovae, gas cooling, and feedback from active galactic nuclei.  These effects can be modeled in hydrodynamical simulations, but differences among the various implementations lead to a spread of predictions implying a theoretical uncertainty in the small scale matter power spectrum~\cite{White:2004kv,Zhan:2004wq,Jing:2005gm,Rudd:2007zx,Semboloni:2011fe,Natarajan:2014xba,Copeland:2019bho,Schneider:2019xpf,Chung:2019bsk,McCarthy:2020dgq,McCarthy:2021lfp}.  Given the vastly different scales and the uncertain physics involved in these processes, it is unlikely that improved simulations alone will entirely remove this uncertainty.  Observational data should provide some insight for baryonic feedback modeling.  However, this poses a challenge for inferring the neutrino mass from measurements of the small scale nonlinear power spectrum, since the effect of neutrino mass is a percent-level nearly scale-independent suppression of power on all but the largest scales.

\subsection{Sensitivity to Neutrino Mass}
\label{subsec:mnu_sensitivity}

\begin{figure}[t!]
    \centering
    \includegraphics[width=5in]{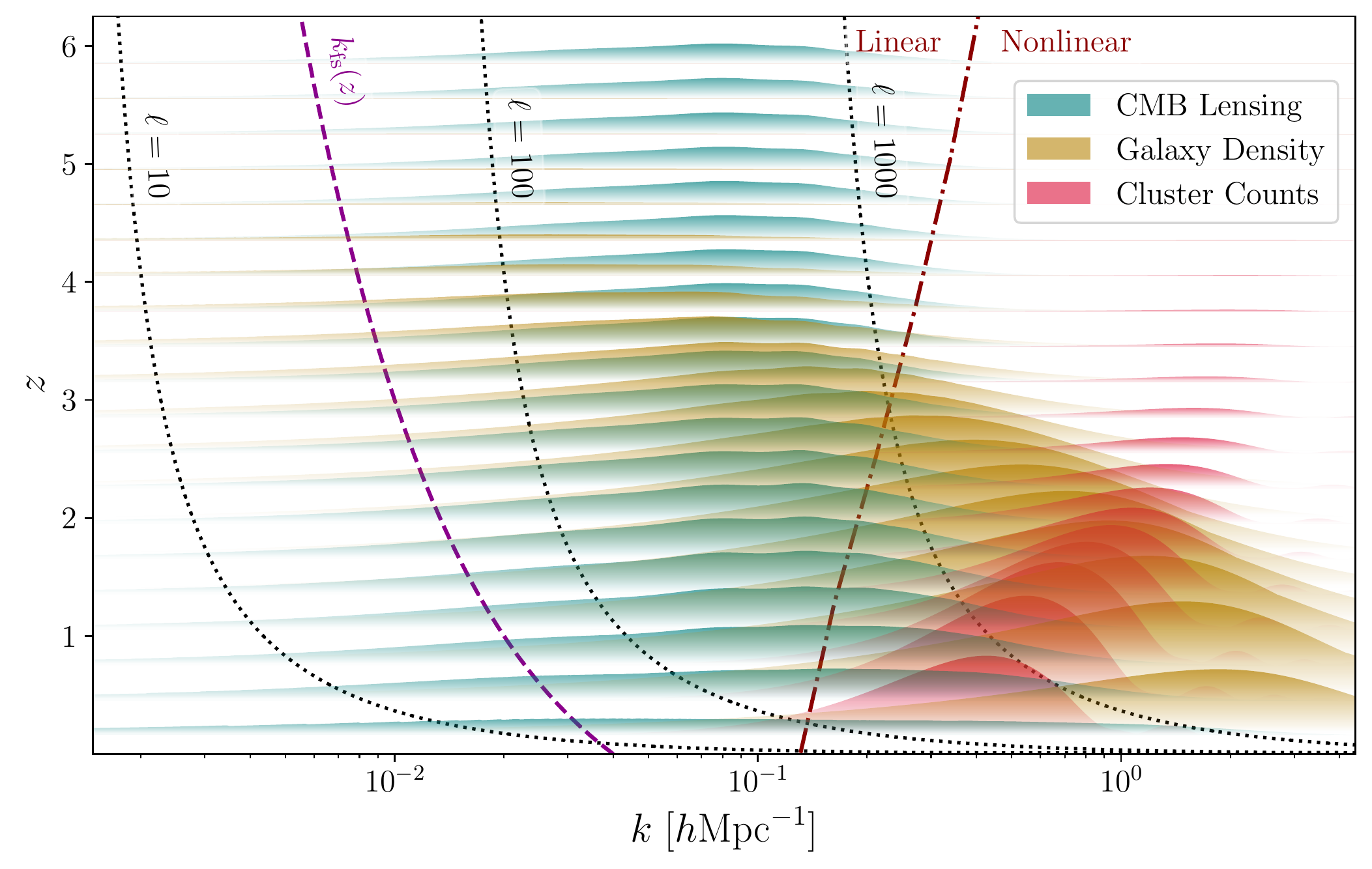}
    \caption{Contributions of the matter power spectrum $P(k,z)$ weighted by signal to noise to the CMB lensing power spectrum with CMB-S4, the angular power of galaxy density with LSST, and number counts of clusters with mass greater than $10^{14}~h^{-1}M_\odot$.  The CMB lensing weighting is scaled by a factor of 3 relative to the others in order to make the CMB lensing contributions more visible despite the very broad lensing redshift kernel.  Black dotted lines denote the values of wavenumber $k$ and redshift $z$ that contribute to a given angular scale $\ell$ in the Limber approximation.  The free-streaming scale $k_\mathrm{fs}(z)$ from Equation~(\ref{eq:kfs_nu}) for standard neutrinos with $\sum m_\nu = 58~\mathrm{meV}$ is shown by the purple dashed line.  Nonlinear corrections to the matter power spectrum are expected to be non-negligible to the right of the red dash-dot line. }
    \label{fig:Pk_stagger}
\end{figure}

In Figure~\ref{fig:Pk_stagger}, we show how the CMB lensing power spectrum $C_\ell^{\kappa \kappa}$, the angular power spectra of galaxy density $C_\ell^{g_i g_i}$, and number counts of galaxy clusters $N_i$ depend upon the matter power spectrum $P(k,z)$ at various wavenumbers $k$ and redshifts $z$.  We weight the contributions of $P(k,z)$ to each observable by its signal-to-noise ratio, using CMB-S4 for CMB lensing~\cite{Abazajian:2019eic}, LSST for galaxy density~\cite{LSSTScience:2009jmu}, and counts of clusters with mass greater than $10^{14}~h^{-1}M_\odot$ corresponding roughly to the detection threshold from the thermal SZ effect as observed by CMB-S4.  

As can be seen from Figure~\ref{fig:Pk_stagger}, the observed CMB lensing power spectrum is dominated by contributions from linear scales across a wide range of redshifts.  Cluster counts are primarily sensitive to the nonlinear matter power spectrum.  The clustering of galaxies at high redshift is sensitive to the matter power spectrum on large scales in the linear regime, while at lower redshifts the signal to noise is dominated by clustering in the nonlinear regime.
This makes CMB lensing a particularly useful probe of neutrino mass, since it provides a measurement of the matter power spectrum on scales where nonlinear corrections and baryonic effects are less significant.  Furthermore, CMB lensing is an unbiased tracer of the matter distribution allowing for direct inferences of the amplitude of the matter power spectrum without marginalizing over bias parameters.

\begin{figure}[t!]
    \centering
    \includegraphics[width=5in]{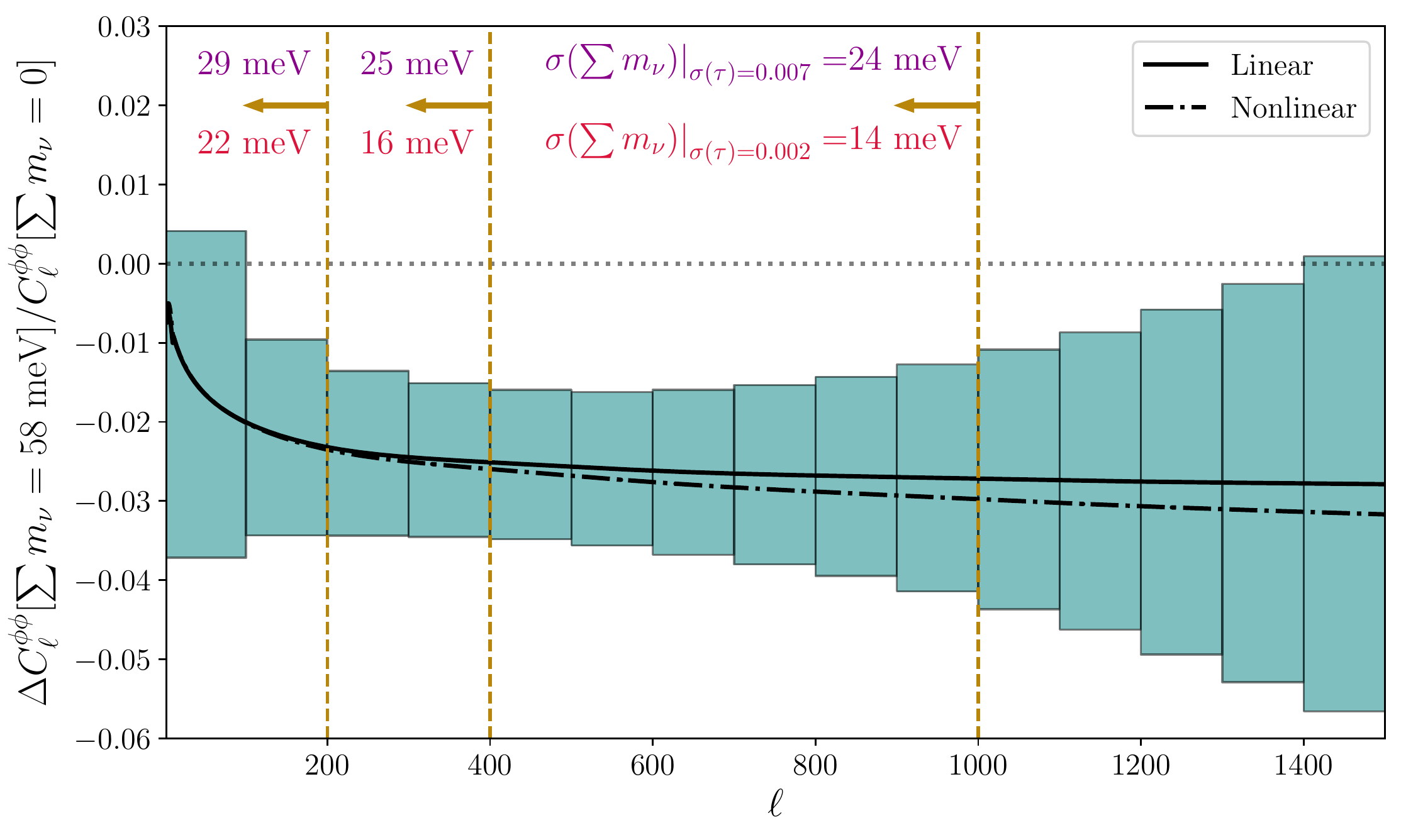}
    \caption{Effect of neutrino mass on CMB lensing power spectrum calculated with the linear matter power spectrum (solid) and the nonlinear matter power spectrum (dash-dot) along with the binned uncertainty expected from CMB-S4.  Forecasted constraints on $\sum m_\nu$ when using CMB lensing information only up to $\ell_\mathrm{max}^{\phi\phi}=[200,400,1000]$ are shown for CMB-S4 + DESI BAO + Planck $\tau$ (top, purple) and for CMB-S4 + DESI BAO + cosmic variance limited $\tau$ (bottom, red).}
    \label{fig:lensing_mnu}
\end{figure}

The impact of neutrino mass on the CMB lensing power spectrum is shown in Figure~\ref{fig:lensing_mnu}.  We also show the binned uncertainty from CMB-S4~\cite{Abazajian:2019eic}, including contributions from both cosmic variance and lensing reconstruction noise.  The suppression of power is apparent across a wide range of angular scales and is nearly constant for $\ell>100$.  The shape-dependence resulting from the imprint of the free-streaming scale appears on large angular scales in the CMB lensing power spectrum where cosmic variance dominates the uncertainty.


We also consider whether nonlinear effects in the matter power spectrum will act as a source of theoretical uncertainty for measurements of $\sum m_\nu$ with CMB lensing.  In Figure~\ref{fig:lensing_mnu} we plot the CMB lensing power spectrum computed using both the linear and nonlinear matter power spectrum, the latter using \texttt{HALOFIT}~\cite{Smith:2002dz,Takahashi:2012em,Bird:2011rb}.  We present forecasts for $1\sigma$ constraints on $\sum m_\nu$ using CMB-S4 along with the DESI BAO~\cite{Font-Ribera:2013rwa} and for two measurements of the optical depth $\tau$: one using the current uncertainty from Planck $\sigma(\tau) = 0.007$~\cite{Planck:2018vyg}, and another assuming a future cosmic variance limited measurement $\sigma(\tau) = 0.002$.  We present forecasts using a set of cuts on the small scale CMB lensing information.  As can be seen from the forecasted constraints, due to the limitations imposed by the uncertainty on our measurement of the optical depth, the vast majority of the $\sum m_\nu$ constraining power from CMB lensing comes from large angular scales $\ell<400$.  The effect of nonlinear corrections to the matter power spectrum is totally negligible on these large scales, implying that modeling of the nonlinear matter power spectrum will have only a very small impact on our ability to infer $\sum m_\nu$ from measurements of CMB lensing.  Similarly, baryonic effects on the CMB lensing power spectrum are most prominent on small angular scales~\cite{Chung:2019bsk}, and the need for their precise modeling could be obviated by restricting analysis of CMB lensing to large scales without much loss in our ability to measure $\sum m_\nu$~\cite{McCarthy:2020dgq}. 

Turning this around, a measurement of $\sum m_\nu$ would allow us to more precisely constrain the amplitude of small scale power predicted by linear theory. A measurement of $\sum m_\nu = 58$ meV at $5\sigma$ with each of these probes requires control of the amplitude of the power spectrum at the sub-percent level.  CMB lensing is expected to provide a reliable measure of the amplitude of the matter power spectrum on linear scales, thus making the consistency of galaxy clustering and cluster counts a sub-percent test of nonlinear structure formation.  Combined with galaxy surveys and cluster counts, this could be used to probe baryonic feedback mechanisms and structure formation on nonlinear scales.

In addition, a variety of extensions of the Standard Model suppress power on small scales.  One well-studied example is dark matter-baryon interactions~\cite{Dvorkin:2013cea,Gluscevic:2017ywp,dePutter:2018xte,Gluscevic:2019yal}, where small scale clustering of dark matter is suppressed through efficient scattering with the photon-baryon fluid prior to recombination.  Large scale structure constraints on these types of scenarios weaken when these interactions involve only a sub-component of the dark matter and may be degenerate with the neutrino mass signal in a given probe.  However, by combing probes, any change to the amplitude of clustering between the scales probed by CMB lensing and the scale of galaxies and clusters would be a window into a wide range of models that alter the matter power spectrum on short distances~\cite{Gluscevic:2019yal}.

A dark energy equation of state $w$ greater than that of a cosmological constant $w=-1$ can also lead to a suppression of power on small scales due to its effects on cosmic distances and on the growth of structure~\cite{Frieman:2008sn}.  The amplitude of suppression as a function of redshift differs for dark energy and neutrino mass; however, since CMB lensing is a measure of the matter power spectrum integrated over redshift, there exists a degeneracy between $\sum m_\nu$ and $w$ (and also the redshift dependence of the equation of state, $w_a$) as measured by CMB lensing~\cite{Allison:2015qca}.  Galaxy clustering, galaxy lensing, and cluster counts (as well as their cross correlations with CMB lensing) can help to break this degeneracy since they provide sensitivity to the growth of structure as a function of redshift~\cite{Madhavacheril:2017onh,Schmittfull:2017ffw,Yu:2018tem,Abazajian:2019eic,Yu:2021vce,Raghunathan:2021zfi}.

\section{Dark Sectors}
\label{sec:dark}

The search for dark matter covers 90 orders of magnitude in mass and coupling strengths from strong to gravitational.  We do not know if the dark matter is composed of a single species or if it is part of a complicated dark sector that includes a mixture of hot, cold, and warm components.  These possibilities are highly motivated by solutions~\cite{Graham:2015cka,Chacko:2016hvu,Chacko:2018vss} to the hierarchy problem and/or the strong CP~\cite{Peccei:1977hh, Weinberg:1977ma, Hook:2018dlk} problem which also predict new particles with sub-eV masses.

Our ability to detect cosmic neutrinos is a remarkable demonstration of the power of cosmology to determine properties of particles from their gravitational influence alone.  This makes cosmology an excellent complement to direct and indirect searches for new particles that rely on specific couplings to the Standard Model~\cite{Gluscevic:2019yal}.

One challenge for cosmology is to make the connection between the particle physics and cosmological observables so that a complete re-analysis of cosmological data is not required for every new idea.  Instead, we would like to be able to reinterpret constraints on parameters of a single ``simplified model" as capturing the essential physics of a much wider class of possibilities.  This philosophy has been applied to great effect in colliders~\cite{LHCNewPhysicsWorkingGroup:2011mji,Buchmueller:2013dya,Cheung:2013dua,Cohen:2013xda} and dark matter searches~\cite{Abdallah:2015ter}. 

In this section, we will show the combination of cosmological constraints on $\sum m_\nu$ and $\Neff$ can serve as a useful simplified model for a wide range of dark sectors.  For essentially massless particles, $m \ll 50$ meV, this subject has been well-explored in the context of $\Neff$ alone~\cite{Green:2019glg}.  However, for larger masses, the relationship between $\sum m_\nu$ constraints and the details of the dark sector models have been less transparent.  Our goal in this section is to show how a detection of $\sum m_\nu$ can be directly interpreted as a constraint on dark sector physics.

\subsection{Hot Sub-Components}


Let us first consider the possibility of a single stable relic, $\chi$, with mass $m_\chi$, number density $n_\chi$, and temperature $T_\chi$ at redshift zero.  We also assume that there exists cold dark matter that is decoupled from $\chi$.  We would like to understand how, and under what circumstances, the cosmological influences of $\chi$ can be mapped to $\Neff$ and $\sum m_\nu$. 

First, one can impose the constraint that the energy density does not exceed the total matter density in the universe
\beq
m_\chi n_\chi < \rho_\mathrm{m} \ .
\eeq
We are interested in particles that act as hot matter, so they would have been relativistic during Big Bang nucleoysnthesis (BBN), such that 
\beq
\Delta \Neff^{\rm BBN} = \frac{4}{7} \, g_{\chi,\star} \left(\frac{T_\chi}{T_\nu}\right)^4 \ ,
\eeq
where $g_{\chi,\star}$ is the effective number of spin degrees of freedom of $\chi$, including an additional factor of $7/8$ for fermionic particles. 
Recent limits on the primordial element abundances gives the constraint $\Delta \Neff^{\rm BBN} < 0.37$ at 95\%~\cite{Cyburt:2015mya}.  

To understand the limits on this model from CMB and late time surveys, we should then determine the redshift $z_\chi$ where $\chi$ becomes non-relativistic such that $\langle p_\chi \rangle / m_\chi <1$.  Assuming the particle is free-streaming, we can write this as
\beq
1 + z_\chi =  \frac{m_\chi}{3 T_\chi} \ .
\eeq
All the modes observed in the CMB $(\ell <4000)$  entered the horizon at $z < 10^5$
and therefore any particle with $z_\chi > 10^5$ will be indistinguishable from cold dark matter as far as the CMB is concerned (unless it is heated at low redshifts).

The presence of a non-zero mass will appear as a change to $\sum m_\nu$ when the wavenumber corresponding to the free-streaming scale of $\chi$ is comparable to or smaller than that of neutrinos, $k_{\rm fs}^{\chi} \lesssim k^{\nu}_{\rm fs}$.  Since the energy density in the non-relativistic regime is $\rho_\chi = m_\chi n_\chi$, the effective sum of neutrino masses is then
\beq\label{eq:mnueff}
    \sum m_{\nu,{\rm eff}} = \sum m_\nu + g_{\chi} m_\chi \left(\frac{T_\chi}{T_\nu}\right)^3 = \sum m_\nu \left(1 +  \frac{1+ z_\chi}{1+z_{\nu}} h_\chi \Delta \Neff^{\rm BBN} \right) \ .
\eeq
where $g_\chi$ is the effective number of degrees of freedom for the number density, including a $3/4$ factor for fermionic particles, and $h_\chi = g_\chi / g_{\chi, \star} \approx 1$.

\begin{figure}[ht!]
    \centering
    \includegraphics[width=5in]{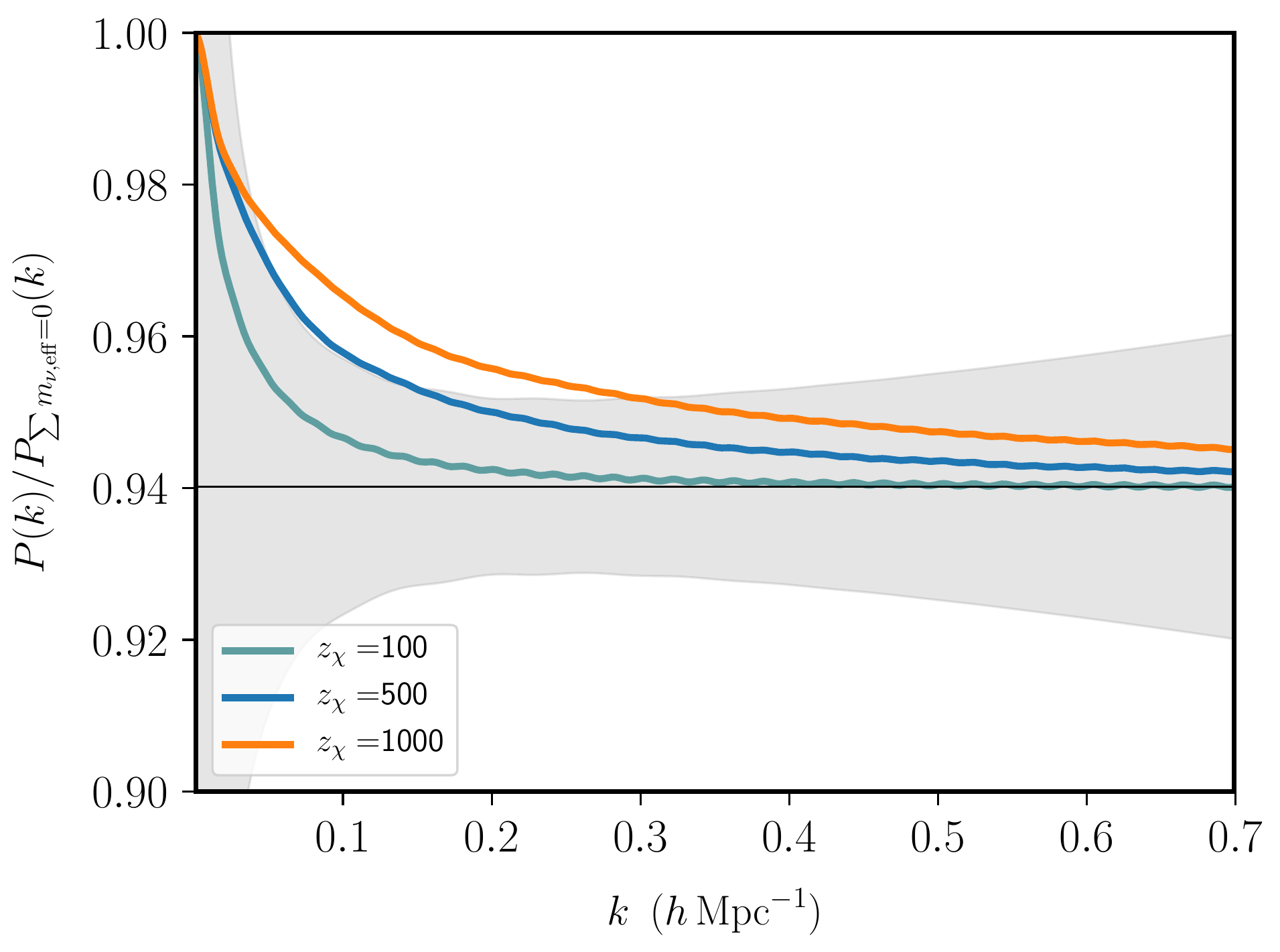}
    \caption{Power spectrum suppression at $z=0$ for various values of $z_\chi$ with fixed values for $\sum m_{\nu, {\rm eff}} = 120$~meV and $\sum m_\nu = 58$ meV.  Here we see the impact on $k_{\rm fs}$ at $z=0$ compared to the 1$\sigma$ variance from BOSS in grey.  This figure shows that the cosmological effects of new species with $z_\chi \lesssim 1000$ should be reliably captured by changes to $\sum m_\nu$ and $\Neff^{\rm CMB}$.
    }
    \label{fig:zchi}
\end{figure}

This description is only approximate, of course, as the impact of $\chi$ on the matter power spectrum is affected by the redshift where $\chi$ becomes non-relativistic.  First, the amplitude of the effect is logarithmically sensitive to $z_\chi$, by comparing with the suppression in Equation~\eqref{eq:suppression}.  Second, it changes the free-streaming scale to
 \beq\label{eq:kfs_chi}
    k^{\chi}_{\rm fs}(z) = \sqrt{\frac{3}{2}}\frac{aH}{c_\chi} = k^{\nu}_{\rm fs}(z) \, \left( \frac{m_\chi}{\sum m_\nu}\right) \, \left( \frac{T_\nu}{T_\chi} \right) =  k^{\nu}_{\rm fs}(z) \, \frac{1+ z_\chi}{1+z_{\nu}}
 \eeq
where we used $z_\chi = 3 T_\chi /m_\chi$.  The impact of changing $z_\chi$ is shown in Figure~\ref{fig:zchi}. We see that for $z_\chi > 1000$, the scale dependence of the signal would be visible on observable scales in current large-scale structure surveys.  In contrast, if $z_\chi \leq 1000$, $k_{\rm fs}$ is too small to be resolved in near-term surveys and the signal is degenerate with $\sum m_\nu$.

Finally, if $z_\chi < 1100$, the relic becomes non-relativistic after recombination.  In this regime, the energy density prior to recombination contributes to $\Neff$ as
\beq
    \Delta \Neff^{\rm CMB} = \frac{4}{7} \, g_{\chi,\star} \left(\frac{T_\chi}{T_\nu}\right)^4
\eeq
where $g_{\chi,\star}$ is the effective number of degrees of freedom of our additional field.  This effect is present even when $m_\chi = 0$.  

\subsubsection*{Example 1: Thermal Relics}

First let us consider light particles that were in thermal equilibrium at very high temperatures but decoupled from the Standard Model at some temperature $T_\mathrm{F}$.  In thermal equilibrium the number density of these light particles is determined solely by the number of internal degrees for freedom.  After the temperature drops below $T_\mathrm{F}$, this number density is diluted relative to the photons only by the number of Standard Model degrees of freedom that annihilated below $T_\mathrm{F}$ and thus gives universal predictions for $\Delta\Neff$ (see e.g.~\cite{Green:2019glg} for discussion).  One finds that
\beq
\Delta \Neff^{\rm BBN}  \geq   0.027 \, g_{\chi,\star} \ ,
\eeq
where the inequality is saturated for $T_\mathrm{F}$ above the electroweak scale.  If we assume this bound is saturated, then
\beq
T_\chi = 0.47 \, T_\nu \, , \qquad 1 + z_\chi = 4200 \times \left(\frac{m_\chi}{1\, {\rm eV}}\right) \ .
\eeq
Note that even in the minimal case, the temperature is still nearly the temperature of the neutrinos, and therefore the dependence of $T_\chi$ on $T_\mathrm{F}$ is very weak. 

\begin{figure}[ht!]
    \centering
    \includegraphics[width=5in]{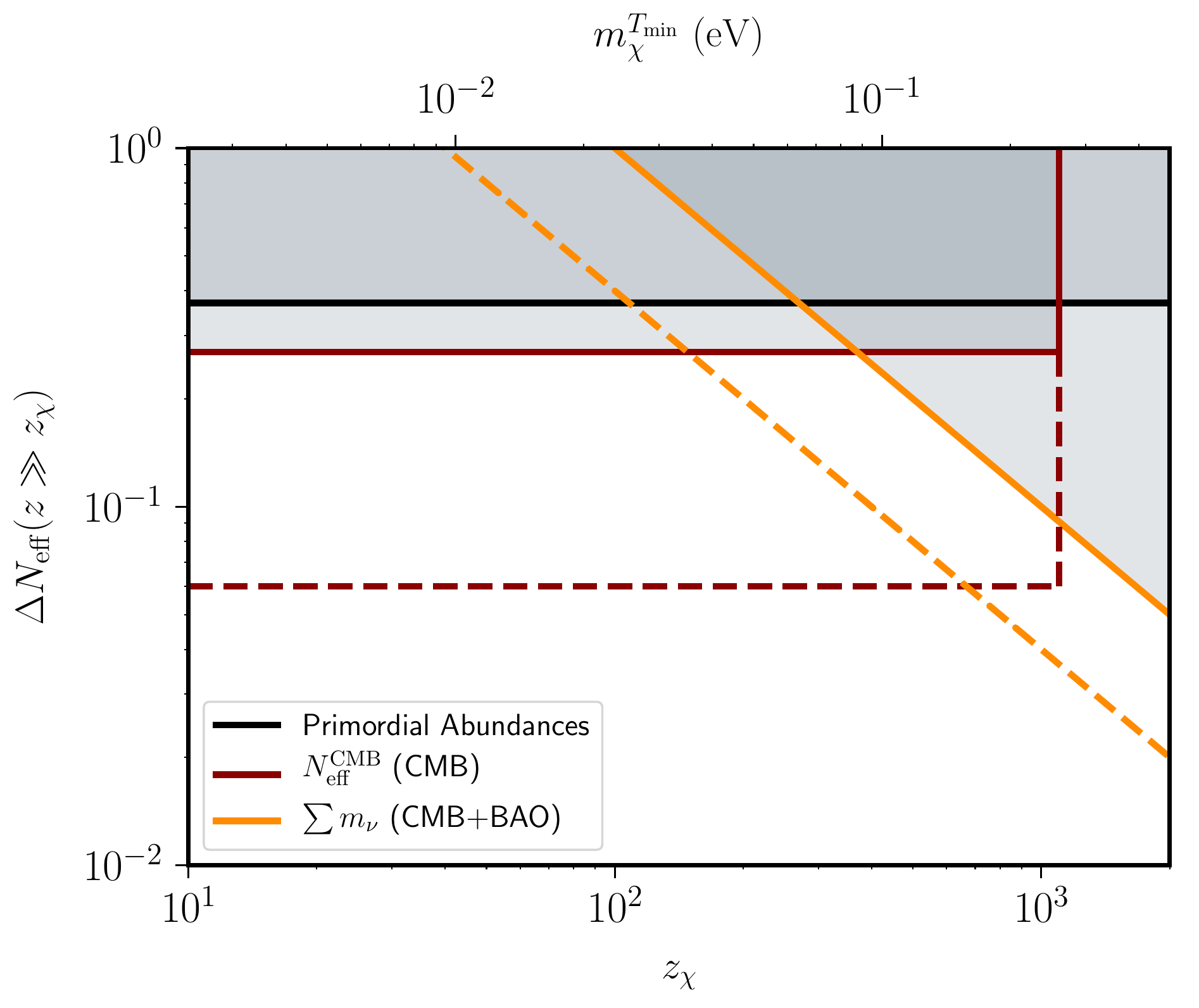}
    \caption{Exclusions of dark sectors in terms of $\Delta \Neff$ and $z_\chi$.  We limited the range to $z_\chi < 2000$ where we expect the neutrino mass constraint to apply directly. The horizontal axis is labeled on top with the values of $m_\chi$ that correspond to values of $z_\chi$ assuming the minimum $T_{\chi}$ from the thermal freeze-out, $T_\chi = 0.47 \, T_\nu$.  The solid lines indicate current constraints from primordial abundances~\cite{Cyburt:2015mya} and Planck~\cite{Planck:2018vyg} with existing BAO measurements~\cite{Beutler:2011hx,Ross:2014qpa,BOSS:2016wmc}, while the dashed lines are projections for CMB-S4~\cite{Abazajian:2019eic} with DESI BAO~\cite{Font-Ribera:2013rwa}.
    }
    \label{fig:Neff_zchi}
\end{figure}

The constraints on these models from measurements of $\Neff$ and $\sum m_\nu$ are shown in Figure~\ref{fig:Neff_zchi}.  With current observations, $\sum m_\nu$ provides a constraint on dark sectors with masses ${\cal O}(100\, {\rm meV})$ and with enough degrees of freedom to reach $\Neff \sim 0.1$, just below current limits.  We also see clearly the complementarity between measurement of $\Neff$ and $\sum m_\nu$ in this parameter space.

For masses larger than an eV, a dedicated analysis is required, as $k_{\rm fs}$ is on observable scales~\cite{DePorzio:2020wcz}.  Current constraints from BOSS~\cite{Xu:2021rwg} exclude masses greater than 11, 2.3, 1.1~eV for a single massive scalar, Weyl fermion, or vector, respectively.  The Weyl fermion case is particularly interesting as a constraint on massive gravitinos from low energy supersymmetry breaking.  Similar constraints can be derived from the CMB, particularly with the addition of CMB lensing data~\cite{Osato:2016ixc}.


\subsubsection*{Example 2: Hidden Sectors}

Next suppose we have a complex hidden sector with many degrees of freedom but with a temperature $T_\chi$ that is not determined by equilibrium with the Standard Model.  Given the constraint from $\Neff^{\rm BBN}$, we conclude that $T_\chi \ll T_\nu$.  Since the contribution of light species in the dark sector to $\Neff$ is proportional to $T_\chi^4$, we might expect to have $\Delta\Neff \ll 1$ unless $T_\chi$ was very finely tuned.  This possibility is well-motivated from solutions to the hierarchy problem~\cite{Arkani-Hamed:2016rle,Chacko:2016hvu,Craig:2016lyx} and from top-down model building~\cite{Kolb:1985bf,Green:2007gs,Feng:2008mu}.  

For simplicity, let us assume this dark sector has a typical mass scale $m_\chi$.  From Equation~(\ref{eq:mnueff}), if $\Delta\Neff \ll 1$ then the neutrino mass signature will also be highly suppressed except when $z_\chi \gg z_\nu$.  However, from Equation~(\ref{eq:kfs_chi}) we see that this corresponds precisely to taking the free-streaming scale $k_{\rm fs}^{\chi} \gg k_{\rm fs}^\nu$.  As a result, when $\Neff$ is parametrically small and far below current or future limits, the signal of the hidden sector will either be too small to observe or will be indistinguishable from cold dark matter.  As a result, we see in Figure~\ref{fig:Neff_zchi} that $\sum m_\nu$ is primarily a valuable probe in the regime where $\Delta\Neff$ is comparable to the observational sensitivity.

\subsection{Dark Sectors Interactions}


A remarkable feature of a cosmological detection of the minimum neutrino mass is that such a measurement would indicate that $0.4\%$ of the matter in the universe has left a signal observed at high significance.  This implies we are sensitive to components of the universe that make up only $0.1\%$  of  the matter density, purely through their gravitational effects.  

Naturally, this suggests that a similarly small fraction of the dark matter with exotic properties could be identifiable from its impact on the late universe.  A simple illustration is provided by a scenario in which a fraction of the matter, $f_\chi$, acquires a velocity $c_\chi \approx c_\nu$ for redshifts $z < z_{c_\chi}$, perhaps by scattering efficiently with some source of radiation.  In this case, we would have a second free-streaming component of matter with 
\beq
    k_{{\rm fs}, \chi} =  \sqrt{\frac{3}{2}}\frac{aH}{c_\chi(z)} \, ,  \qquad \omega_\chi = f_\chi \omega_\mathrm{m} \ .
\eeq
The redshift dependence of $c_\chi(z)$ depends on the precise mechanism that imparts the velocity.  Nevertheless, when $k \gg k_{{\rm fs},\chi}$, we the matter power spectrum will be suppressed by 
\beq
    P_{f_\chi} (k\gg k_{{\rm fs},\chi}, z ) \approx  \left(1 - 2 f_\chi  -\frac{6 }{5} f_\chi\log \frac{1+z_{c_\chi}}{1+z} \right) P_{f_\chi =0}(k\gg k_{{\rm fs},\chi}, z ) \, . \label{eq:xi_suppression}
\eeq
Provided $k_{{\rm fs},\chi}\ll 0.1 \, h \, {\rm Mpc}^{-1}$, $f_\chi \neq 0$ will appear as a shift to $\sum m_\nu$.  However, since we know $\sum m_\nu \geq 58$ meV, a high significance detection of the minimum neutrino mass would place severe constraints on $f_\chi$.  Concretely for plausible future measurements with CMB-S4+DESI of $\sum m_\nu = 58 \pm 13$ meV~\cite{Abazajian:2019eic}, we would constrain $f_\chi < 0.002$ at 95\% (assuming $z_{c_\chi} = {\cal O}(100)$).

Note that, in contrast to hot relics, the scale $z_{c_\chi}$ is not related to $k_{\rm fs,\chi}$.  Instead, $z_{c_\chi}$ defines the redshift when the dark matter first acquires its large velocity.  A concrete way to achieve a large velocity is to couple a fraction of the matter, $f_\chi$, to a source of radiation with a cross-section that increases at low-redshifts.  A simple realization of this idea was found in Ref.~\cite{Green:2021gdc}, where the dark matter was coupled via a light mediator to right-handed neutrinos (see also~\cite{Wilkinson:2014ksa,Archidiacono:2014nda,Binder:2016pnr,Boehm:2017dze,DiValentino:2017oaw,Stadler:2019dii,Becker:2020hzj,Mosbech:2020ahp}).  For a large range of parameters, the dark matter scatters efficiently with the neutrinos for $z < 1100$ and produces a signal degenerate with a change to the neutrino mass.  The same signal could be achieved by coupling directly to photons~\cite{Dvorkin:2013cea,Gluscevic:2017ywp,dePutter:2018xte,Gluscevic:2019yal} (e.g. milli-charged dark matter) or to some hidden sector radiation, although these may be subject to additional constraints from the CMB~\cite{Cyr-Racine:2013jua,Park:2019ibn,Brinckmann:2020bcn}, and from direct and indirect dark matter searches.

\section{Discussion}
\label{sec:Discussion}

The cosmological measurement of neutrino mass is among the most highly anticipated products of the next generation of cosmological surveys.  Cosmological measurements currently provide the most stringent constraints on $\sum m_\nu$ and are expected to present the first determination of the absolute neutrino mass scale.  A detection consistent with the minimum mass implied by neutrino flavor oscillations, $\sum m_\nu = 58$~meV, would both constitute a stringent test of our cosmological models and confirmation that these cosmic neutrinos are the same neutrinos we observe directly on Earth or indirectly through their role in stellar evolution.  

In this paper, we explored how the campaign to detect the effects of neutrino mass in cosmic surveys will inform our understanding of the universe.  These measurements are of broad cosmological value for two keys reasons.  First, they provide a remarkably sensitive test, measuring the unique properties of a sub-percent component of the matter in the universe.  As a result, they place severe constraints on any other sub-percent component of the energy density that would similarly alter the matter power spectrum.  Second, a measurement at this level of sensitivity is expected from a variety of cosmological probes that are sensitive to different redshifts and distance scales.  Consistent measurements of $\sum m_\nu$ with these different observables informs our knowledge of the dynamics of the universe on these disparate scales.  

We have focused on the near-term measurement of neutrino mass, which manifests itself as a suppression of power in lower redshift observables.  However, in principle, the neutrino mass introduces other unique cosmological signatures that may also be tested in the future.  The most obvious target would be to directly observe the scale $k_{\rm fs}$ that indicates on onset of free-streaming.  In addition, the large amplitude of the suppression of small scale power stems from the change to the growth rate of matter and thus the neutrino mass could also be measured through redshift dependence of the growth~\cite{Yu:2018tem}.
This signal could be seen via CMB secondaries, and would help to bolster confidence in a detection by other means due to the different systematic effects that impact the measurements~\cite{Madhavacheril:2017onh}. 
Additional effects arising from the nonlinear formation of structure such as scale-dependent bias~\cite{LoVerde:2016ahu}, relative velocity of neutrinos and dark matter~\cite{Zhu:2013tma}, and neutrino wakes~\cite{Zhu:2014qma} provide additional signatures that are not present in the absence of neutrino mass.  While these additional measurements are beyond the scope of our discussion, they would directly probe the neutrino mass in such a way that we could separate $\sum m_\nu$ from other effects that suppress small scale power.  As we have explained, the detection of $\sum m_\nu$ in the near term will be degenerate with a wide range of physical effects expected in $\Lambda$CDM and beyond, and these additional signals will remain important targets even in the presence of a cosmological detection of a non-zero neutrino mass.

\vskip20pt
\paragraph{Acknowledgments}

We thank Shaul Hanany for valuable discussions and for encouraging us to publish these results. We also thank Nathaniel Craig, Cora Dvorkin, Raphael Flauger, Yi Guo, David~E.~Kaplan, Lloyd Knox, Marilena LoVerde, Surjeet Rajendran, and Ben Wallisch for helpful discussions. DG is supported by the US~Department of Energy under Grants~\mbox{DE-SC0019035} and~\mbox{DE-SC0009919}.  JM is supported by the US~Department of Energy under Grant~\mbox{DE-SC0010129}. We acknowledge the use of \texttt{CLASS}~\cite{Blas:2011rf}, \texttt{IPython}~\cite{Perez:2007ipy}, and the Python packages \texttt{Matplotlib}~\cite{Hunter:2007mat}, \texttt{NumPy}~\cite{Harris:2020xlr}, and~\texttt{SciPy}~\cite{Virtanen:2019joe}.

\appendix

\section{CMB Lensing Reconstruction}
\label{app:LensingRecon}

In this appendix, we briefly review how gravitational lensing changes the statistics of the CMB anisotropies, and how those effects can be used to reconstruct a map of the lensing potential.  Using the methods described below, the low-noise and high-resolution measurements of the CMB from upcoming surveys~\cite{Abazajian:2016yjj,Ade:2018sbj,Hanany:2019lle,Sehgal:2019ewc} will provide precise measurements of CMB lensing.  As described in Section~\ref{sec:clustering} the CMB lensing power spectrum is a powerful tool for measuring the matter power spectrum, and in particular is sensitive to the suppression of clustering imprinted by neutrino mass.

The lensed CMB temperature is given in terms of the unlensed temperature and the lensing deflection as
\begin{equation}
    T^\mathrm{len}(\hat{\mathbf{n}}) = T(\hat{\mathbf{n}}+\boldsymbol{\alpha}(\hat{\mathbf{n}})) \, ,
    \label{eq:lensedT}
\end{equation}
and similarly the Stokes $Q$ and $U$ describing CMB polarization are lensed according to
\begin{align}
    Q^\mathrm{len}(\hat{\mathbf{n}}) &= Q(\hat{\mathbf{n}}+\boldsymbol{\alpha}(\hat{\mathbf{n}})) \, , \nonumber \\
    U^\mathrm{len}(\hat{\mathbf{n}}) &= U(\hat{\mathbf{n}}+\boldsymbol{\alpha}(\hat{\mathbf{n}})) \, .
    \label{eq:lensedQU}
\end{align}
The $E$- and $B$-mode polarization fields are related to $Q$ and $U$ in the flat-sky approximation (valid on small angular scales) by
\begin{align}
    \left[Q\pm iU\right](\hat{\mathbf{n}}) = - \int \frac{\dd^2\ell}{(2\pi)^2}\left[E(\boldsymbol{\ell}) \pm i B(\boldsymbol{\ell}) \right] e^{\pm i 2 \varphi_{\boldsymbol{\ell}} } e^{i \boldsymbol{\ell} \cdot \hat{\mathbf{n}}} \, ,
    \label{eq:EB_def}
\end{align}
where $\hat{\mathbf{x}}\cdot \hat{\boldsymbol{\ell}} = \cos \varphi_{\boldsymbol{\ell}}$.
Lensing deflection results in the correlation between the observed CMB fluctuations with different wavenumber.
For example, the two-point correlation function of the lensed $E$ and $B$ averaged over CMB fluctuations but with fixed lensing deflection is given by~\cite{Hu:2001kj}
\begin{align}
    \left\langle E(\boldsymbol{\ell}_1) B(\boldsymbol{\ell}_2) \right\rangle_{\mathrm{CMB}}
    = f_{EB}(\boldsymbol{\ell}_1,\boldsymbol{\ell}_2) \phi(\mathbf{L})  \, ,
    \label{eq:Lensing_EB}
\end{align}
where the $EB$ mode-coupling is
\begin{align}
    f_{EB}(\boldsymbol{\ell}_1,\boldsymbol{\ell}_2) = 
    \left[C_{\ell_1}^{EE}(\mathbf{L} \cdot \boldsymbol{\ell}_1) -
    C_{\ell_2}^{BB}(\mathbf{L} \cdot \boldsymbol{\ell}_2) \right] 
    \sin 2(\varphi_{\boldsymbol{\ell}_1} - \varphi_{\boldsymbol{\ell}_2}) \, ,
    \label{eq:EB_Mode_Coupling}
\end{align}
and $\mathbf{L} = \boldsymbol{\ell}_1 + \boldsymbol{\ell}_2 \neq 0$.

This induced correlation can be used to estimate the lensing potential by computing a weighted sum of a product of $E$ and $B$ modes
\begin{align}
    \hat{\phi}_{EB}(\mathbf{L}) = A_{EB}(\mathbf{L}) 
    \int \frac{\dd^2\ell_1}{(2\pi)^2} E(\boldsymbol{\ell}_1) B(\boldsymbol{\ell}_2) F_{EB}(\boldsymbol{\ell}_1, \boldsymbol{\ell}_2) \, ,
    \label{eq:EB_estimator}
\end{align}
where $\boldsymbol{\ell}_2 = \mathbf{L}-\boldsymbol{\ell}_1$, and $A(\mathbf{L})$ is a normalization defined as
\begin{align}
    A_{EB}(\mathbf{L}) = \left[ \int \frac{\dd^2\ell_1}{(2\pi)^2} f_{EB}(\boldsymbol{\ell}_1, \boldsymbol{\ell}_2) F_{EB}(\boldsymbol{\ell}_1, \boldsymbol{\ell}_2) \right]^{-1} \, ,
    \label{eq:Normalization}
\end{align}
chosen to make the estimator unbiased
\begin{align}
    \left\langle \hat{\phi}_{EB}(\mathbf{L}) \right\rangle_\mathrm{CMB} = \phi(\mathbf{L}) \, .
    \label{eq:Unbiased}
\end{align}
The filter $F_{EB}(\boldsymbol{\ell}_1, \boldsymbol{\ell}_2)$ is chosen to minimize the variance of the estimator and is given by
\begin{align}
    F_{EB}(\boldsymbol{\ell}_1, \boldsymbol{\ell}_2) = \frac{ f_{EB}(\boldsymbol{\ell}_1, \boldsymbol{\ell}_2) }{C_{\ell_1}^{EE,\mathrm{obs}} C_{\ell_2}^{BB,\mathrm{obs}} } \, ,
    \label{eq:EB_Filter}
\end{align}
where the observed spectra include the effects of lensing as well as instrumental noise (plus any residual systematic effects or foregrounds) $C_\ell^{EE,\mathrm{obs}} = C_\ell^{EE,\mathrm{len}} + N_\ell^{EE}$.
In the case of white noise, the noise power is 
\begin{align}
    N_\ell^{EE} = N_\ell^{BB} = \Delta_P^2 \exp \left( \ell(\ell+1) \frac{\theta_\mathrm{FWHM}^2}{8\ln 2} \right) \, ,
    \label{eq:Noise_Power}
\end{align}
where $\Delta_P$ is the instrumental polarization noise in units of $\mu$K-rad, and $\theta_\mathrm{FWHM}$ is the full-width at half-maximum beam size in radians.
The total variance of the estimator is given by
\begin{align}
    \left\langle \hat{\phi}_{EB}(\mathbf{L}) \hat{\phi}_{EB}(\mathbf{L}') \right\rangle = 
    (2\pi)^2 \delta(\mathbf{L} + \mathbf{L}') \left[C_L^{\phi\phi} + N_{EB}(\mathbf{L}) \right] \, ,
    \label{eq:EB_Variance}
\end{align}
and with the choice of minimum variance filter given in Equation~\eqref{eq:EB_Filter}, we have $N_{EB}(\mathbf{L}) = A_{EB}(\mathbf{L})$.

Similar estimators can be constructed for each of the six pairings of the fields $T$, $E$, and $B$.
At the low noise expected from future CMB surveys, the $EB$ lensing estimator will have the lowest variance due to the small observed $B$-mode power.
The lensing reconstruction noise can be reduced by estimating and removing the lensing-induced $B$-mode polarization.
The improved lensing estimate can be used to remove more lensing-induced $B$-mode power, and this procedure can be iterated until the lensing reconstruction noise converges, giving an optimal estimate of the lensing map~\cite{Hirata:2003ka,Smith:2010gu}.

\phantomsection
\addcontentsline{toc}{section}{References}
\small
\bibliographystyle{utphys}
\bibliography{nu}

\end{document}